\documentclass[12pt]{article}

\usepackage{graphicx}
\usepackage{a4}

\usepackage{fancybox}
\usepackage{epsfig}
\usepackage{color}

\usepackage{amsfonts}
\usepackage{mathrsfs}
\usepackage{amssymb}
\usepackage{amsmath}

\usepackage{dcolumn}
\usepackage{bm}

\def\pa{\partial}

\def\be{\beta}

\def\de{\delta}
\def\ep{\epsilon}

\newcommand{\ben}{\begin{equation}}
\newcommand{\een}{\end{equation}}
\newcommand{\bea}{\begin{eqnarray}}
\newcommand{\eea}{\end{eqnarray}}
\newcommand{\ba}{\begin{array}}
\newcommand{\ea}{\end{array}}
\newcommand{\bit}{\begin{itemize}}
\newcommand{\eit}{\end{itemize}}

\newcommand{\vs}[1]{\vspace{#1 mm}}

\newcommand{\dsl}{\pa \kern-0.5em /}

\begin{document}

\topmargin 0pt \oddsidemargin 0mm




\vspace{2mm}

\begin{center}

{\Large

\bf Can matter really cross a horizon?}

\vs{10}

 {\large Huiquan Li \footnote{E-mail: lhq@ynao.ac.cn}} 

\vspace{6mm}

{\em

Yunnan Observatories, Chinese Academy of Sciences, \\
650011 Kunming, China

Key Laboratory for the Structure and Evolution of Celestial Objects,
\\ Chinese Academy of Sciences, 650011 Kunming, China}

\end{center}

\vs{9}

\begin{abstract}
It has been taken as a truth that collapsing matter can eventually
cross the horizon and enter into the interior of a black hole in a
finite proper time. However, the Rindler/tachyon dual description we
suggest recently implies that this should not be the case. A test
particle falling towards the event horizon of a non-extreme black
hole can actually be viewed as an unstable particle, whose dynamics
is described by the tachyon field theory. This means that the
collapsing process of a free particle in Rindler space is
essentially a tachyon condensation process. In terms of the results
in tachyon condensation, we learn that the infalling particle should
strongly couple to bulk modes and should decay completely into
something like gravitons before reaching the horizon. Hence, there
should be no matter that can cross a horizon as still matter. The
matter will get ``dissolved" into spacetime when approaching the
event horizon.
\end{abstract}



\begin{center}

\center{\Large Essay written for the Gravity Research Foundation
2014 Awards for Essays on Gravitation}

\center{Submission date: 31 March, 2014}

\end{center}


\newpage

It has become a well-known fact that material reaching the event
horizon will be inevitably sucked into the black hole. In other
words, collapsing material can cross the horizon and get inside the
black hole in a finite proper time, as analysed in classical and
semi-classical methods. In such kind of treatments of the physical
processes near a non-extreme black hole, we usually do not need to
worry about the thermal bath environment and other quantum effects.
However, we indeed should be very careful in dealing with physics in
this region. In this essay, we shall point out that a quantum effect
is ignored in these treatments, that is, the coupling between the
collapsing material and the background gravitational field. This
will be clearly seen by the aid of the tachyon field theory, which
is well studied in string theory.

Tachyon is originally referred to a particle moving faster than
light. In modern physics, in particular in string theory, it has a
different meaning: tachyon field is a scalar field with negative
mass squared, but moving no faster than light. The imaginary mass
means that the field can behave in growing or decreasing modes
rather than the oscillating modes. Its appearance signals
instability of the system it describes. The unstable process of the
system can be described by the tachyon field rolling down the
asymptotic potential, i.e., the tachyon condensation process. A
couple of different effective theories, which are believed to be
equivalent, have been developed for the tachyon field. Among them,
the boundary conformal field theory (BCFT) and the tachyon effective
field theory are two popular ones (see \cite{Sen:2004nf} for a
review). Some key results about tachyon condensation are derived in
these two theories. In string theory, tachyon field is found to
exist as the lowest state of the exciting spectrum of unstable
strings and branes. In what follows, we shall show that tachyon is
also related with physics near non-extreme black holes
\cite{Li:2011ypa,Li:2013wwa}.

The near-horizon geometry of non-extreme black holes generically
contains the Rindler space:
\begin{equation}\label{e:}
 ds_R^2=e^{-2\be T}(-dt^2+dT^2)=-\rho^2dt^2+\frac{1}{\be^2}d\rho^2,
\end{equation}
with
\begin{equation}\label{e:}
 T=-\frac{1}{\be}\ln\rho.
\end{equation}
It is well known that the Hawking temperature due to the periodicity
of the Euclidianlised $t$ coordinate is:
\begin{equation}\label{e:Hawtemp}
 T_{\textrm{Haw}}^{(t)}=\frac{\be}{2\pi}.
\end{equation}
In the universal time coordinate, the physics near the horizon
becomes ``freezing" because the time evolves very slowly. As
observed at spatial infinity, it takes infinite time for an
infalling particle to reach the horizon. But it spends finite time
for the particle to cross the horizon and to reach the singular
center of the black hole, as analysed in standard textbooks.

The dynamics of a freely infalling particle (or point-like object)
with mass $m_0$ in the above Rindler space is described by the
following effective action:
\begin{equation}\label{e:tacact}
 S_{0}=-\int d t V(T)\sqrt{1-\dot{T}^2},
\end{equation}
where
\begin{equation}\label{e:}
 V(T)=2\tau_0e^{-\be T}, \textrm{ }\textrm{ }\textrm{ }
\tau_0=\frac{m_0}{2}.
\end{equation}
Approaching the event horizon $\rho\rightarrow0$, we have
$T\rightarrow\infty$ and $V(T)\rightarrow0$. This action is exactly
the tachyon effective action \cite{Garousi:2000tr,Kutasov:2003er}
with the full potential $V(T)=\tau_0/\cosh(\be T)$ as the tachyon
field $T$ grows large at late times.

So the infalling particle in Rindler space can be viewed as an
unstable particle. If no energy is lost\footnote{In string theory,
this happens when the string coupling constant tends to zero.}, the
energy-momentum tensor at late times is derived from the action:
\begin{equation}\label{e:enmom}
 T_{00}=\frac{V(T)}{\sqrt{1-\dot{T}^2}}=E,
\textrm{ }\textrm{ }\textrm{ } T_{ij}=-\frac{V^2}{E}\de_{ij}
\simeq-\frac{4\tau_p^2}{E} e^{-2\be t}\de_{ij},
\end{equation}
So the infalling particle or the tachyon reach a pressureless and
freezing state $T_{ij}\rightarrow0$ as $t$ and so $T(t)$ tend to
infinity (i.e., when reaching the horizon). This state corresponds
to the formation of tachyon matter \cite{Sen:2002in}, which is
believed to be massive closed strings radiated during the
condensation \cite{Sen:2003bc,Sen:2004nf}.

The tachyon effective theory has an equivalent description, the
BCFT. It can reproduce the results obtained in BCFT
\cite{Okuyama:2003wm,Lambert:2003zr,Sen:2004nf}, including the
energy-momentum tensor (\ref{e:enmom}). There are two important
results about tachyon condensation has been obtained in the two
theories, with more detailed calculations elaborated in BCFT. In
what follows, we present the main conclusions and their implications
to the collapsing particle.

(1) \textbf{Particle (open-string) pair creation from rolling
tachyon} This has been discussed in
\cite{Strominger:2002pc,Gutperle:2003xf,Maloney:2003ck}. It is
derived that the tachyon creates particle pairs at the Hagedorn
temperature (in the time coordinate $t$)
\begin{equation}\label{e:Hagtemp}
 T_{\textrm{Hag}}^{(t)}=\frac{\be}{2\pi}.
\end{equation}
It is easy to notice that this temperature is the same as the
Hawking temperature (\ref{e:Hawtemp}), detected in the same time
coordinate.

The Hagedorn temperature usually indicates that the system reaches
the critical temperature for phase transition. So we can say that
the particle collapsing process towards the event horizon of a
non-extreme black hole is also a Hagedorn phase transition process,
with the Hawking temperature being the Hagedorn temperature. Thus,
the infalling particle should become something else.

(2) \textbf{Graviton (closed-string) radiation from rolling tachyon}
The rolling tachyon can couple to the closed strings represented by
$\Psi$ \cite{Lambert:2003zr}
\begin{equation}\label{e:}
 S=\frac{1}{\ep}\int dtL[T(t)]+\int dtL[T(t)]\Psi(t,x=0)
-\frac{1}{2}\int dtd^qx[(\pa\Psi)^2+m^2\Psi^2],
\end{equation}
where $L[T]=-V\sqrt{1-{\dot{T}}^2}$. When the string coupling
constant (proportional to $\ep$) is not small, the coupling to
massive closed strings is exponentially strong and the tachyon will
dump its energy into closed-string radiation quickly
\cite{Okuda:2002yd}. In \cite{Okuda:2002yd,Lambert:2003zr}, detailed
calculations show that an unstable particle will completely decay
into closed strings, mainly very massive ones, towards the end of
condensation.

This means that the infalling particle which is described by the
effective theory (\ref{e:tacact}) will completely decay into
gravitons (or closed strings) before reaching the horizon. There are
two cases. (a) Massless graviton (closed-string) emission: In string
theory, Hawking radiation can be interpreted as massless
closed-string radiation on the horizon \cite{Callan:1996dv}. So the
massless graviton emission from the infalling particle is suppressed
because it is in thermal equilibrium with the Hawking radiation,
since the Hawking and Hagedorn temperatures are equal. (b) Thus,
most energy of the particle is transferred into the massive graviton
emission, forming pressureless tachyon matter, as found in
\cite{Lambert:2003zr}. The degenerated states of the emitted
gravitons from the infalling particle should be responsible for the
increased entropy in the black hole due to absorption of the
particle. It has been checked in \cite{Li:2011ypa} that the result
is comparable with the conjectured value, in particular for black
holes far away from extremality.

Based on the above analysis and results, we may further speculate
that there should exist a dual description between physics near
non-extreme black holes and tachyon field theory. The correspondence
between two sides are summarised in Table.\ \ref{t:t1}.
\begin{table}
\begin{center}
\begin{tabular}{|c|c|}
\hline
\textbf{(Thermo)dynamics in Rindler space} & \textbf{Tachyon field theory} \\
\hline
The event horizon & The closed string vacuum \\
\hline
Freezing physics near the event horizon & Freezing state of tachyon matter \\
\hline
Hawking temperature & Hagedorn temperature \\
\hline
Increased area $\de A/4$ & Entropy $\de S$ of degenerate states of  \\
due to absorption of a particle & the closed strings that the particle decays into \\
\hline
\end{tabular}
\caption{\label{t:t1} Correspondence between the two sides of the
Rindler/tachyon dual description.}
\end{center}
\end{table}
From this, we may have the conclusion that \textit{an infalling
particle can not cross a horizon as still matter, because it will
act as an unstable particle near the event horizon and will decay
completely into something like gravitons before reaching the
horizon}.


\bibliographystyle{JHEP}
\bibliography{b}

\end{document}